# Distributed Parametric Effect in Long Lines and its Applications

Yuri K. Shestopaloff

The article considers a parametric effect which takes place when the velocity of signal propagation in a long line changes. We found the analytical solution describing the form of the transformed signal for a line with losses, when line parameters change symmetrically. We also considered lines without losses, with asymmetrical change of parameters. Our theoretical results comply with experimental data. In certain conditions, such a line can be used as an amplifier. The parametric effect in optics is described by Maxwell's equations, while in case of a long line, the analysis is based on telegrapher's equations. However, it turns out that in the end, both in optics and electronics, the parametric effect is described by wave equations that are mathematically similar. This is because fundamentally, the parametric effect is physically based on energy interchange between the controlling (pump) signal and the transformed one, when the parameters of the propagating medium change. So, the obtained results can be used for analysis of parametric effects in optics and electronics.

## Introduction

The parametric effect is used in different applications, such as in parametric amplifiers in optics [1] and electronics, in particular in high frequency range of electromagnetic emission [2]. In essence, the main principle of a parametric amplifier is to transfer energy of controlling signal, which is used to change parameters of some circuit elements, such as, for instance, capacitors, to the signal to be amplified. The controlling signal is often called a pump signal. In optics, such a pump signal changes the optical parameters of a fiber line.

The parametric effect can be realized in discrete elements, such as capacitors, or be distributed along the propagating medium. Physically, the amplification effect can be explained as follows. Let us assume that the controlling signal changes parameters of a discrete circuit element in the presence of the useful signal. The system composed of the useful signal and the circuit element, understood as one entity, changes under the influence of the controlling signal. The controlling signal changes parameters of the element *and* useful signal's characteristics simultaneously; at this moment the useful signal is inherently associated with the element, both constitute one entity, and so it is impossible to change the element's parameters without changing the characteristics of the useful signal. When the useful signal is amplified, the source of energy is the controlling signal. The reverse, that is amplification of controlling signal by the energy of the useful signal, is also possible, although this effect is rarely mentioned.

We consider a distributed parametric system, namely a long line. The word "long" in this context essentially means that the "electrical" length of the line is greater than the wavelength. In other words, what matters is the relation of the propagation time and duration of signal. So, we can also define the long line as the one in which the time of signal propagation is greater than or at least comparable to the duration of signal.

There is one more aspect associated with the propagation of a signal in long lines with changing parameters. The change of line's parameters may cause the change of signal's velocity of propagation, which accordingly changes the characteristics of this signal. For instance, if we pass a rectangular impulse through a line in which the propagation velocity increases, then the duration of impulse will be reduced. This effect was noticed a long time ago. One of the earliest applications of this effect is described in [3]. In this work, the authors suggest to divide a continuous signal into impulses that are passed through a long line with increasing propagation velocity. As a result, such a line produces an output of separate impulses instead of a continuous signal. Disregarding the input between impulses, the authors suggested that this way the overall noise of a receiver, which uses this effect, can be reduced. However, the authors assumed that such a line is passive, which is not true. In fact, the signals in the long line are amplified because of the presence of the parametric effect, so that the actual noise reduction is less than the authors claimed.

Although we consider long lines, most distributed parametric amplifiers are adequately described by the wave equation in one form or another [1, 4], whether in the optical or microwave range of wavelengths. So, many results we obtain below can be transferred to other ranges of the electromagnetic spectrum.

## 1. Finding the output signal in the long line without losses

The velocity of propagation $v$ of a signal in a long line is defined as follows.

$$v = 1/\sqrt{LC} \tag{1}$$

where $L$ and $C$ are accordingly the distributed inductance and capacity of the line.

We assume that the velocity of propagation changes in such a way that the signal $f(t)$, where $t$ is time, undergoes a linear transformation $P$, i.e. $P(f(t)) = rf(\theta \times t)$ for some constants $r$ and $\theta$. In particular, this is true when the velocity changes as follows [5].

$$v = \frac{v_0}{1+bt} \tag{2}$$

where $b$ is constant; $v_0 = 1/\sqrt{L_0 C_0}$ ; $L_0$ and $C_0$ are accordingly the initial distributed inductance and capacity.

So, in order to change the propagation velocity, we have to accordingly change the distributed inductance or capacity, or both. The velocity of propagation and accordingly line parameters begin to change at the moment when the signal appears in the line and change until the signal leaves the line. Sometimes, this type of velocity change is called simultaneous control of line parameters. Alternatively, it is possible to create a line in which each value of a signal at a given moment will be transferred through the line at a constant speed, although the velocity will be different for different parts of signal. For simultaneous control, the authors in [4] found that the input signal $f(t)$ in the line without losses transforms into

$$u(x,t) = \exp\left(-\frac{bx}{v_0}\right) \times f\left[t \times \exp\left(-\frac{bx}{v_0}\right) + \frac{1}{b}\left(\exp\left(-\frac{bx}{v_0}\right) - 1\right)\right] \quad (3)$$

Here, $u(x,t)$ is an output voltage, $x$ denoting distance from the beginning of the line. Propagation velocity changes according to (2), which is done by changing the distributed inductance and capacity by a controlling signal as follows.

$$C = C_0(1+bt) \quad (4)$$

$$L = L_0(1+bt) \quad (5)$$

When $b > 0$, the signal stretches, because the velocity of propagation decreases and the back part of signal propagates slower than the signal front. If $b < 0$, then the signal, accordingly, compresses. Authors of [4] did not consider change of signal's energy. However, as we noted earlier, the energy changes as well. We can show this as follows. Let us assume that the parameters of line change symmetrically according to (4) and (5). In this case, its characteristic impedance $Z$ is constant, that is

$$Z = \sqrt{\frac{L}{C}} = \sqrt{\frac{L_0(1+bt)}{C_0(1+bt)}} = \sqrt{\frac{L_0}{C_0}} = Z_0 \quad (6)$$

Then, the current at the line's input is defined as follows: $I(x,t) = \dfrac{u(x,t)}{Z_0}$. Assuming that the duration of input signal is $T$, we can find the signal energy at line input as follows.

$$E_0 = \int_0^T I(x,t) \times u(x,t) dt \quad (7)$$

Using (3), we find that the duration of signal at the output is defined as $(t_2 - t_1)$, where

$$t_1 = -\frac{1}{bk}(k-1); \quad t_2 = \frac{T}{k} - \frac{1}{bk}(k-1); \quad k = \exp\left(-\frac{bx}{v_0}\right).$$

Then, the energy of the output signal is as follows.

$$E = \int_{t_1}^{t_2} k^2 I(kt + \frac{1}{b}(k-1)) \times u(kt + \frac{1}{b}(k-1)) dt \tag{8}$$

Substituting in (4) $p = kt - \frac{1}{b}(k-1)$, we find

$$E = k \int_0^T I(p) \times u(p) dp = kE_0 \tag{9}$$

We can see from (9) that compressed signal increases its energy, while the energy of stretched signal reduces. This is due to energy interchange between the controlling signal and input signal. Works [3, 4] did not reveal the active character of long line with changing parameters. Authors of [3] suggested dividing the continuous signal into separate parts and passing them through the long line, thus compressing each part of the signal, so that the signal parts will be separated at the output of the long line. Figure 1 presents the idea. Disregarding the interval in which the signal is absent, the authors could potentially reduce the overall receiver noise. However, the assumption about passive character of such line led to overestimation of noise reduction.

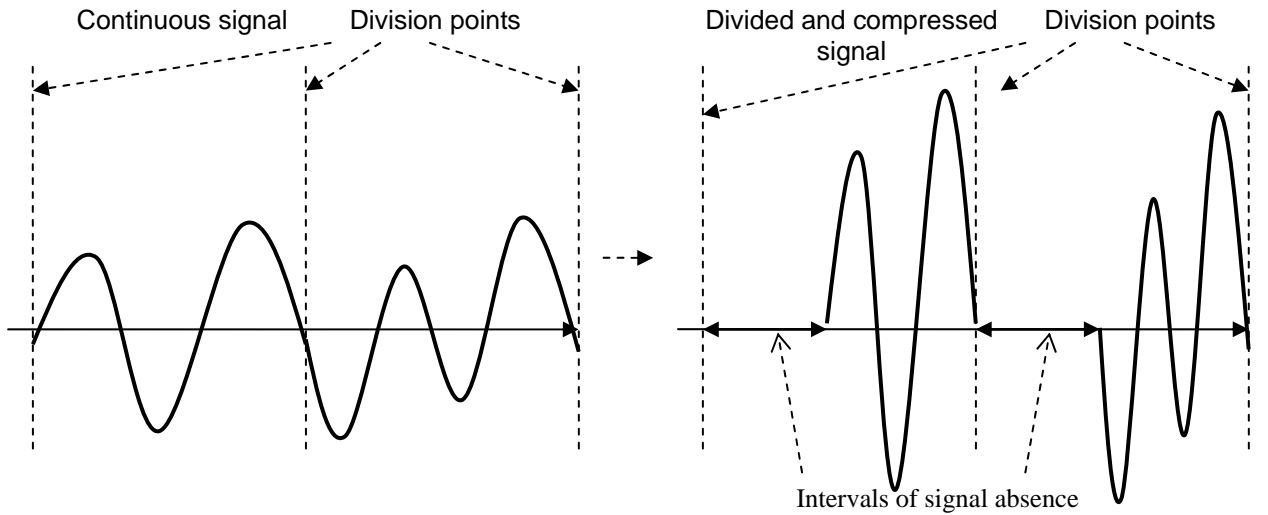

Figure 1. Continuous signal is divided and compressed by the long line.

## 2. Change of spectrum of transformed signals

In this section, we consider change of spectrum of transformed signal that is passed through the long line with changing propagation velocity. It is reasonable to assume that the spectrum of compressed signal should increase. The spectrum of a signal is defined as follows.

$$g(\omega) = \int_{-\infty}^{+\infty} \psi(\tau) \exp(-i\omega\tau) d\tau = \int_{-\infty}^{+\infty} \overline{f(t)f(t+\tau)} \exp(-i\omega\tau) d\tau \qquad (10)$$

Recall that we consider the linear transformation $f(kt)$ of the signal. Accordingly, the spectrum of the transformed signal can be found similarly to (10) as follows.

$$g_t(\omega) = \int_{-\infty}^{+\infty} \overline{f(kt)f(kt+k\tau)} \exp(-i\omega\tau) d\tau =$$

$$= \frac{1}{k} \int_{-\infty}^{+\infty} \overline{f(kt)f(kt+k\tau)} \exp\left(-i\left(\frac{\omega}{k}\right)k\tau\right) d(k\tau) = \frac{1}{k} g\left(\frac{\omega}{k}\right) \qquad (11)$$

It follows from (11) that the bandwidth of compressed signal increases, while the bandwidth of the stretched signal decreases. This is another factor that should be taken into account in distributed parametric amplifiers.

## 3. Transformation of signal in long line with losses

If the long line has losses, then the telegrapher's equations for such a line are as follows.

$$\frac{\partial u}{\partial x} + L\frac{\partial i}{\partial t} + i\frac{\partial L}{\partial t} + R_0 i = 0 \qquad (12)$$

$$\frac{\partial i}{\partial x} + C\frac{\partial u}{\partial t} + u\frac{\partial C}{\partial t} + g_0 u = 0 \qquad (13)$$

where $g_0$ and $R_0$ are accordingly distributed conductivity and resistance; $i$ is current.

There are different methods for solving telegraph equations depending on their particular form. Although we found interesting approaches suggested in [6-8] and some other works, we directly find an analytical solution. Let us consider a symmetrical change of distributed capacity and inductance according to (4) and (5). Then, (12) and (13) can be transformed to the following equation.

$$\frac{\partial^2 u}{\partial x^2} - LC\frac{\partial^2 u}{\partial t^2} - (1+bt)a\frac{\partial u}{\partial t} - qu = 0 \qquad (14)$$

where $a = 3L_0C_0 b + L_0 g_0 + R_0 C_0$; $q = (R_0 + L_0 b)(g_0 + bC_0)$.

In order to perform this transformation, we find second derivatives of (12) with respect to $x$ and (13) with respect to $t$ and then substitute $\frac{\partial i}{\partial x}$ from (13) and $\frac{\partial^2 i}{\partial x \partial t}$ from the second derivative of (13) into the second derivative of (12).

We can do the substitution $\tau = \frac{v_0}{b}\ln(1+bt)$. Then, (14) can be transformed to

$$\frac{\partial^2 u}{\partial x^2} - \frac{\partial^2 u}{\partial \tau^2} + d\frac{\partial u}{\partial \tau} - qu = 0 \qquad (15)$$

Here, $d = b/v_0 - av_0$.

We can get rid of the first derivative using the following substitution: $u = w \times \exp\left(\frac{1}{2}d\tau\right)$. Then, (15) will change as follows.

$$\frac{\partial^2 w}{\partial x^2} - \frac{\partial^2 u}{\partial \tau^2} + pw = 0 \qquad (16)$$

where $p = \frac{1}{4}d^2 - q$.

Let the input signal be $f(t)$. Then, we have the following boundary conditions for (14):

$u(0,t) = f(t)$; $\frac{\partial u(0,t)}{\partial x} = 0$. For variables $x$ and $\tau$ the same conditions are as follows.

$$u(0,\tau)_{x=0} = f\left[\frac{\exp(b\tau/v_0)-1}{b}\right]$$

$$w(x,\tau)_{x=0} = f\left[\frac{\exp(b\tau/v_0)-1}{b}\right] \times \exp\left(-\frac{1}{2}d\tau\right) = f_1(\tau)$$

If we introduce new variables $\xi$ and $\eta$ such that $\tau = \dfrac{\xi + \eta}{2}$, $x = \dfrac{\xi - \eta}{2}$, then we obtain the canonical form of equation (16).

$$\frac{\partial^2 w}{\partial \xi \partial \eta} + mw = 0 \tag{17}$$

where $m = -p/4$.

The condition $x = 0$ corresponds to $\xi = \eta$. Hence, we have:

$$\left.\frac{\partial w}{\partial \xi}\right|_{\eta=\xi} = \frac{1}{2} f_1^{(1)}(\xi); \quad \left.\frac{\partial w}{\partial \eta}\right|_{\eta=\xi} = \frac{1}{2} f_1^{(1)}(\eta); \quad w(\xi,\eta)_{\eta=\xi} = f_1(\xi) \tag{18}$$

Equation (17) can be solved by Riemann's method. The Riemann function is as follows.

$$R(\xi,\eta,\xi_0,\eta_0) = J_0\left(\sqrt{4m(\xi-\xi_0)(\eta-\eta_0)}\right) \tag{19}$$

Here, $J_0(y)$ is Bessel's function of order zero. Using Riemann function (19), we find the solution as follows.

$$u(x,t) = \frac{1}{2}\left\{ e^{-\frac{bx}{v_0}} e^{-(L_0 g_0 + R_0 C_0) v_0 x} f\left[e^{-\frac{bx}{v_0}} t + \frac{1}{b}\left(e^{-\frac{bx}{v_0}} - 1\right)\right] + e^{\frac{bx}{v_0}} e^{-(L_0 g_0 + R_0 C_0) v_0 x} f\left[e^{\frac{bx}{v_0}} t + \frac{1}{b}\left(e^{\frac{bx}{v_0}} - 1\right)\right]\right\} \tag{20}$$

Formula (20) presents direct and reverse waves. If we assume that the reverse wave is absent (the line impedance matches the output load) then we only have the direct wave.

$$u(x,t) = e^{-\frac{bx}{v_0}} e^{-(L_0 g_0 + R_0 C_0) v_0 x} f\left[e^{-\frac{bx}{v_0}} t + \frac{1}{b}\left(e^{-\frac{bx}{v_0}} - 1\right)\right] \tag{21}$$

Note that the amplitude of the direct wave doubles because we do not have the reverse wave.

Our result differs from the formulas obtained in [4] (long line without losses) by the term $e^{-(L_0 g_0 + R_0 C_0) v_0 x}$. So, the impact of a long line with changing parameters on a signal can be reduced to two transformations of the signal: one is amplification of the signal's power with coefficient *k*, and the other is the signal's compression in the passive line with the loss $e^{2(L_0 g_0 + R_0 C_0) v_0 x}$.

Noise coefficient of line with losses is defined as follows.

$$F = (L-1)\frac{T}{T_0} + 1 \qquad (22)$$

Here, $T$ is thermodynamic temperature of long line (in Kelvin degrees); $T_0 = 290K$.

Once the noise coefficient is known, we can find the noise temperature for a particular line, and accordingly the sensitivity of receiver that uses such signal compression. This is a well known procedure. Note that change of signal spectrum should be taken into account when doing such computations. For instance, if one uses such signal compression to improve the sensitivity of microwave radiometer, then the bandwidth of output signal has to be increased in a way that corresponds to the degree of compression.

## 4. Asymmetrical change of line parameters

Let us consider the case when only the distributed capacity of the line changes, that is $C = C_0(1+bt)^2$. The propagation velocity changes in the same way as in (2), so that this line also performs a linear transformation of the signal. We assume that the long line has no losses. Then, the telegrapher's equations for this line are as follows.

$$\frac{\partial u}{\partial x} + L\frac{\partial i}{\partial t} = 0$$

$$\frac{\partial i}{\partial x} + C\frac{\partial u}{\partial t} + u\frac{\partial C}{\partial t} = 0 \qquad (23)$$

Transforming this system of equations, we obtain the following.

$$\frac{\partial^2 u}{\partial x^2} - \frac{(1+bt)^2}{v_0^2}\frac{\partial^2 u}{\partial t^2} - \frac{4b(1+bt)}{v_0^2}\frac{\partial u}{\partial t} - \frac{2b^2}{v_0^2}u = 0$$

We will do the following substitution: $\tau = \frac{v_0}{b}\ln(1+bt)$. Then, the above equation transforms to the following.

$$\frac{\partial^2 u}{\partial x^2} - \frac{\partial^2 u}{\partial \tau^2} - \frac{3b}{v_0}\frac{\partial u}{\partial t} - \frac{2b^2}{v_0^2}u = 0 \qquad (24)$$

If we substitute $u = w \times \exp\left(\frac{3b\tau}{v_0}\right)$, then (24) can be written as follows.

$$\frac{\partial^2 w}{\partial x^2} - \frac{\partial^2 w}{\partial \tau^2} - \frac{b^2}{4v_0^2} w = 0 \qquad (25)$$

Substituting in (25) $\xi = \tau + x$, and $\eta = \tau - x$, we will obtain the following equation.

$$\frac{\partial^2 w}{\partial \xi \partial \eta} + \frac{b^2}{16 v_0^2} w = 0 \qquad (26)$$

The solution of this equation is similar to equation (17), when we considered a symmetrical change of the line's parameters. We use Riemann's method with the same boundary conditions and assume that the line impedance and the output load dynamically match, so that we do not have the reverse wave. Finally, in the original variables, we obtain the following.

$$u(x,t) = e^{-\frac{3bx}{2v_0}} u_0 \left[ e^{-\frac{bx}{v_0}} t + \frac{1}{b}\left( e^{-\frac{bx}{v_0}} - 1 \right) \right] \qquad (27)$$

Here, $u_0(t)$ is the input signal.

Similarly, solving (23) with respect to current, for the boundary condition $I(x,t)_{x=0} = I_0(t)$, we eventually obtain the following solution.

$$I(x,t) = e^{-\frac{bx}{2v_0}} I_0 \left[ e^{-\frac{bx}{v_0}} t + \frac{1}{b}\left( e^{-\frac{bx}{v_0}} - 1 \right) \right] \qquad (28)$$

If we do transformations similar to (7) – (9) to obtain the overall change of energy, we will find that the signal's energy changes by $k = \exp\left(-\frac{bx}{v_0}\right)$ times. Note that the line's characteristic impedance is changing, so that the input voltage and current change accordingly. However, at any moment their product is equal to the power of signal.

We also considered a long line with constant distributed capacity but varying distributed inductance $L = L_0(1+bt)^2$. The solution can be found in a way similar to (23) – (28) when we analyzed the case of varying capacity and constant inductance. The final result is as follows.

$$u(x,t) = e^{-\frac{bx}{2v_0}} u_0 \left[ e^{-\frac{bx}{v_0}} t + \frac{1}{b}\left( e^{-\frac{bx}{v_0}} - 1 \right) \right]; \quad I(x,t) = e^{-\frac{3bx}{2v_0}} I_0 \left[ e^{-\frac{bx}{v_0}} t + \frac{1}{b}\left( e^{-\frac{bx}{v_0}} - 1 \right) \right]$$

## 5. Experimental verification

The following approach has been used for experimental verification of the discovered parametric effect. The long line was modeled by a delay line with ten varicap diodes and constant inductances uniformly distributed over a length of twenty centimeters. This was done in order to verify the change of signal energy in long lines with varying propagation velocity. Figure 2 presents electrical scheme.

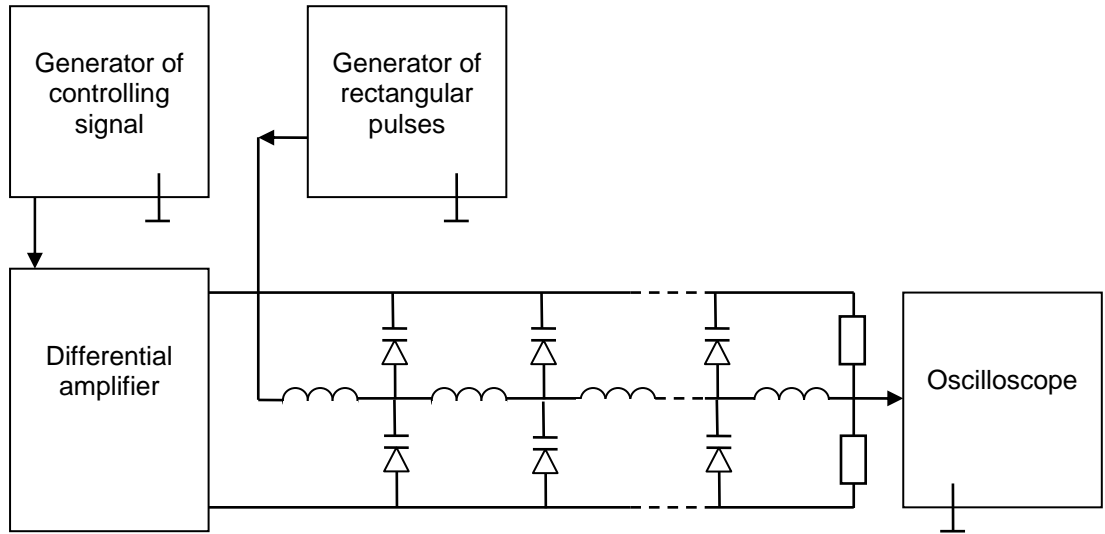

Figure 2. Experimental verification of parametric effect in long lines with distributed parameters.

We used a delay line with ten varicap diodes and constant inductances in order to verify the change of signal energy in long lines with varying propagation velocity. The varicap's capacity changes depending on the reverse voltage $u_r$ as follows: $C = C_0 / \sqrt{u_r}$. In order for the propagation velocity to change as $v = v_0 /(1 + bt)$, the varicaps' controlling voltage has to change as the fourth power of time for signal compression, and inverse of the fourth power of time for signal stretching. Such voltage dependence can be obtained on most standard generators. In order to avoid the impact of a controlling signal, which is sent from a differential amplifier, we used a balance scheme. The duration of controlling signal was 0.3 $\mu s$. The studied signal was a rectangular impulse with duration 0.2 $\mu s$. Imperfect balance of line was eliminated by the following processing. We transmitted positive and negative rectangular signals of the same amplitude through the line for the same operation, let us say for signal compression. Then, during data processing, we performed a symmetrical reflection of one signal relative to abscissa, summed up the signals, and divided the resulting signal by two. As a result of this processing, the influence of misbalancing in the delay line was removed. We used the same procedure to remove the influence of misbalancing for stretched signals. After that, we found the difference in energies between the compressed and stretched signals. We can do this since the impact of controlling signal is the same for stretched and compressed signals. So, subtraction of energies of compressed and stretched signals compensates for the impact of controlling signal.

Table 1. Change of energies of compressed and stretched signals relative to the energy of input signal

| Signal duration | Predicted change of signal energy | Actual change of signal energy |
|---|---|---|
| 0.2 $\mu s$ | 13 % | 15% |

Based on (27), (28), we computed the expected change of signal energy relative to the energy of the input signal of $13\%$. Given the imperfection of controlling signal and distortions of fronts of output signals compared to input signals, the predicted value was in good agreement with the $15\%$ change of signal energy obtained in the experiment, see Table 1. The suggested data processing procedure was crucial for the success of experiment. The additional processing of positive and negative signals allowed us to remove the impact of line unbalancing and the influence of the controlling signal.

Figure 3 shows the power of compressed and stretched signals, while figure 3 shows the difference in power of the same signals.

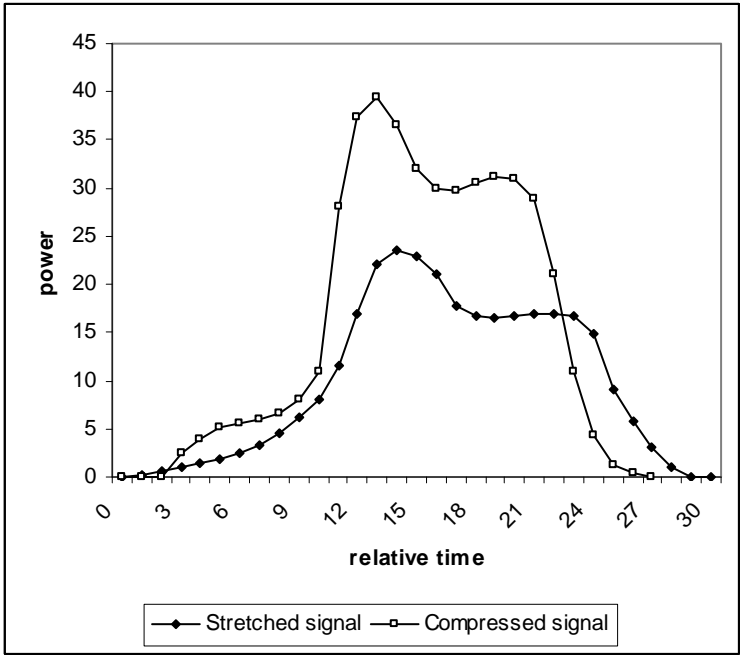

Figure 3. Dependence of power of stretched and compressed signals on time

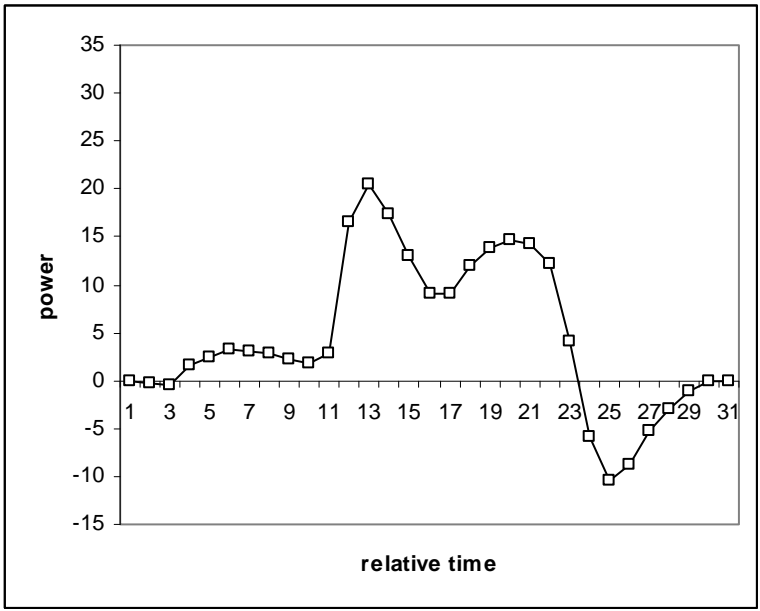

Figure 4. Difference in power of compressed and stretched signals.

## 6. Possible applications

The discovered parametric effect presents several practically meaningful properties. One is the possibility of using the effect for amplification. The other is the ability to change the signal spectrum, that is, to make it wider or narrower. The third possibility is cutting the original, let us say continuous, signal into a sequence of stretched or compressed pulses, composed of the sections of the original signal, and do processing of such pulses in sequence or in parallel. In case of compression, the output signal is similar to the sequence of pulses shown in Figure 1. Then, if such a signal is sent to an amplifier, the amplifier can do amplification only when the signal is present, shutting off between the pulses, which can reduce the overall noise produced by the amplifier. The tradeoff is the necessity to increase the bandwidth, because the spectrum of compressed signal widens. Practical implementation of such an approach is possible, and produces real noise reduction.

The effect can be used for pulse reshaping, for instance, in order to make steeper pulse's fronts. Finally, the effect can also be used for compression, in order to provide resolution or separation of multiple pulses coming into the same location from different directions, for instance, in networks [9].

The obtained results have also general importance with regard to behavior of systems whose parameters change while they interact with another substance, in our case, with electromagnetic field. Apparently, this is a general principle, that is that the energy interchange between the changing system and the entity it interacts with necessarily occurs. We can see the confirmation of this principle in many mathematical formulations of laws of classical mechanics. However, in other areas, the generalization of this energy interchange principle is not so obvious, although, intuitively, it has to have a universal appeal.

**Conclusion**

Work [1] presents solutions for certain types of parametric effects in optics. Mathematically, some solutions are close to what we obtained in our study. The base in both cases is the same – some form of wave equation. From the physical perspective, parametric effects in general are based on the same fundamental mechanism of energy interchange between the controlling and transmitted signals due to the change of parameters of transmitting medium by the controlling signal. This common foundation effectively unites the parametric effects into a single group of physical phenomenon regardless the wavelength range. Thus, the same mathematical apparatus can be used interchangeably across different spectrum ranges. Such a conceptual approach is consistent with the idea of generalization of mathematical methods across several types of functions [10]. Figure 5 illustrates these considerations.

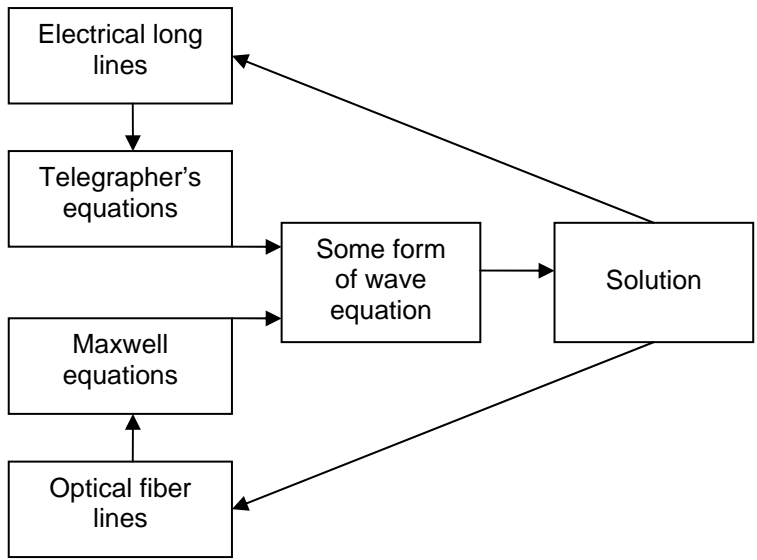

Figure 5. Generalization of solutions of wave equation, applied to the analysis of electrical lines and optical fiber lines.

We would like to emphasize the efficiency of suggested data processing procedures. Similar approaches can be used in other electrical and optical applications to circumvent the signal distortion caused by propagation medium.

Overall, the mathematical model of a long line with losses and symmetrical change of parameters is an adequate way of presenting the parametric phenomenon. The model takes into account all factors meaningful for circuit analysis and design decisions.

We also considered lines without losses and with asymmetrical change of parameters, and consequently with varying impedance. These lines possess a similar property of energy interchange between the controlling and propagating signals. Such a line was implemented in our experimental studies. The experimental results are in good agreement with the theoretical solution.

4. Barna A. (1980), *High speed pulse and digital techniques*, Wiley.

5. Naidenov A. I., Fomin Ch. A. (1968), Transformation of spectrum of electrical signal in the long lines with variable parameters, *Journal of Communication Technology and Electronics*, 23, No. 1.

6. Li Y. (2003), Positive doubly periodic solutions of nonlinear telegraph equations, *Nonlinear Analysis*, 55, 245-254.

7. Bluman G.W., Temuerchaolu (2005), Conservation laws for nonlinear telegraph equation, *Journal of Mathematical Analysis and Applications*, 310, 459-476.

8. El-Azab M.S., El-Gamel M. (2007), A numerical algorithm for the solution of telegraph equations, *Applied Mathematics and Computation*, 190, 757-764.

9. Ninagawa C., Miyazaki Y. (2009), Analysis of Collided Signal Waveform on the Long Transmission Line of UART-CSMA/CD Control Network, *Piers Online*, 5, No. 2, 171.

10. Shestopaloff Yu. K., (2010), *Properties and interrelationships of polynomial, exponential, logarithmic and power functions with applications to modeling natural phenomena*, AKVY Press.
15

**Figure legends**

Figure 1. Continuous signal is divided and compressed by the long line.

Figure 2. Experimental verification of parametric effect in long lines with distributed parameters.

Figure 3. Dependence of power of stretched and compressed signals on time

Figure 4. Difference in power of compressed and stretched signals.

Figure 5. Generalization of solutions of wave equation, applied to the analysis of electrical and optical fiber lines.